\documentclass[aps,twocolumn,groupedaddress,showpacs]{revtex4}
\usepackage[dvips]{graphicx}

\begin{document}


\title{Onsager coefficients of a finite-time Carnot cycle}


\author{Yuki Izumida}
\author{Koji Okuda}
\affiliation{Division of Physics, Hokkaido University, Sapporo 060-0810, Japan}

\begin{abstract}
We study a finite-time Carnot cycle of 
a weakly interacting gas which we can regard as a nearly ideal gas
in the limit of $T_\mathrm{h}-T_\mathrm{c}\to 0$ where $T_\mathrm{h}$ and
$T_\mathrm{c}$ are the temperatures of the hot and cold heat reservoirs,
respectively. 
In this limit, we can assume that the cycle is working
in the linear-response regime and can calculate the Onsager coefficients of this cycle
analytically using the elementary molecular kinetic theory. 
We reveal that these Onsager coefficients satisfy the so-called tight-coupling condition
and this fact explains why the efficiency at the maximal power $\eta_\mathrm{max}$ of this cycle
can attain the Curzon-Ahlborn efficiency from the viewpoint
of the linear-response theory.   
\end{abstract}

\pacs{05.70.Ln}

\maketitle

\section{Introduction}
Improving efficiency of heat engines has been a long challenge
since the Industrial Revolution. Substantial progress of our knowledge 
came from Carnot's insight: He conceived a mathematical model of an
idealized heat engine, now called the Carnot cycle, and showed that 
there is an upper limit of the efficiency of
all the existing heat engines which can be attained only when the
engines are working infinitely slowly (quasistatic limit) to vanish
the irreversibility.

In the fundamental physics, properties of heat engines 
working at the maximal power have also been studied 
since the study by Curzon and Ahlborn~\cite{CA,C} (see also
\cite{N}). 
Although the quasistatic Carnot cycle has the highest efficiency, 
it outputs zero power because it takes infinite time to 
output a finite amount of work. 
By contrast, Curzon and Ahlborn considered a finite-time
Carnot cycle which exchanges heat at a finite rate with the reservoirs 
according to the linear time-independent Fourier law. 
Under the assumption of endoreversibility that irreversible
processes are occurred only through
these heat exchanges, they
derived a remarkable result:
the efficiency at the maximal power $\eta_\mathrm{max}$ is given by 
\begin{eqnarray} 
\eta_\mathrm{max}=1-\sqrt{\frac{T_\mathrm{c}}{T_\mathrm{h}}}\equiv
\eta_\mathrm{CA} \label{eq.1}
\end{eqnarray} 
where $T_\mathrm{h}$ and $T_\mathrm{c}$ are the 
temperatures of the hot and cold heat reservoirs, respectively and the above
$\eta_\mathrm{CA}$
is usually called the Curzon-Ahlborn (CA) efficiency. 

Previously we studied a finite-time Carnot
cycle of a weakly interacting gas which we can regard as a nearly ideal
gas to confirm the validity of the CA efficiency from a more microscopic point of view~\cite{IO,IO2}.
We performed extensive molecular dynamics (MD) computer simulations of the
finite-time Carnot cycle of the two-dimensional low dense hard-disc gas 
and measured the efficiency and the power for the first time. 
Our simulations revealed
that our $\eta_\mathrm{max}$ agrees
with $\eta_\mathrm{CA}$ only in the limit of $\Delta T \to 0$ where
$\Delta T\equiv T_\mathrm{h}-T_\mathrm{c}$, but exceeds
$\eta_\mathrm{CA}$ at somewhat large $\Delta T$. 
We also confirmed this behavior of
$\eta_\mathrm{max}$ analytically using the elementary molecular kinetic theory. 
Therefore the phenomenological prediction of Eq.~(\ref{eq.1}) seems to be valid
only in the limit of the small temperature difference for our finite-time
Carnot cycle. 
Recently, Van den Broeck~\cite{VB} considered the heat engine described
as the Onsager relations 
\begin{eqnarray}
J_1=L_{11}X_1+L_{12}X_2,\label{eq.2}\\
J_2=L_{21}X_1+L_{22}X_2,\label{eq.3}
\end{eqnarray}
and showed that $\eta_\mathrm{CA}$ is the upper limit of
$\eta_\mathrm{max}$ in this heat engine (see also Sec.~V). The recent studies~\cite{GS,SS,TU,ELB1,ELB2} on
the various theoretical heat engine models also support the results in~\cite{VB}.
These results seem to meet our previous result of the
finite-time Carnot cycle, though it is unclear why our system realized 
the upper limit of $\eta_\mathrm{max}$ in $\Delta T \to 0$.
Moreover it is also unclear whether the finite-time Carnot cycle can be
understood in the framework of the Onsager relations
because the explicit calculations of the Onsager
coefficients $L_{ij}$ for that cycle do not exist to our knowledge. 

In this paper, we apply the framework of the Onsager relations to our previous
study of the finite-time Carnot cycle and analytically calculate
the Onsager coefficients for it. 
Although there are a few analytic calculations of
the Onsager coefficients for the steady state of heat
engines such as Brownian motors~\cite{GS,BK}, we believe that  
the present study is the first example of the calculation for the cyclic heat
engine model where the two heat reservoirs do not contact with the working
substance simultaneously. 
We will show that these Onsager coefficients satisfy the
so-called tight-coupling condition and therefore we can give an explicit
explanation why $\eta_\mathrm{max}$ of this cycle attains the CA efficiency in the
limit of $\Delta T \to 0$, as observed in~\cite{IO,IO2}, from the
viewpoint of the linear-response theory. We also perform the MD computer
simulations to check the validity of our analytic calculations of the 
Onsager coefficients.

The organization of this paper is as follows. First we introduce  
our finite-time Carnot cycle model in Sec.~II and describe the molecular
kinetic theory in Sec.~III. 
The main result of this paper, the analytic calculations of the Onsager      
coefficients for our model, accompanied by the results of the MD      
simulations for checking the validity of our analytic calculations, are  
shown in Sec.~IV. In Sec.~V, we introduce the general framework of the heat engine
used in~\cite{VB} and 
discuss the efficiency at the maximal power of our 
finite-time Carnot cycle using those Onsager coefficients according to
that framework.
We summarize this study in Sec.~VI. 
\section{Model}
We first introduce a theoretical model for a finite-time Carnot cycle
of a two-dimensional weakly interacting gas which we can regard as a nearly ideal gas~\cite{IO,IO2}.
\begin{figure}
\begin{center}
\includegraphics[scale=0.3]{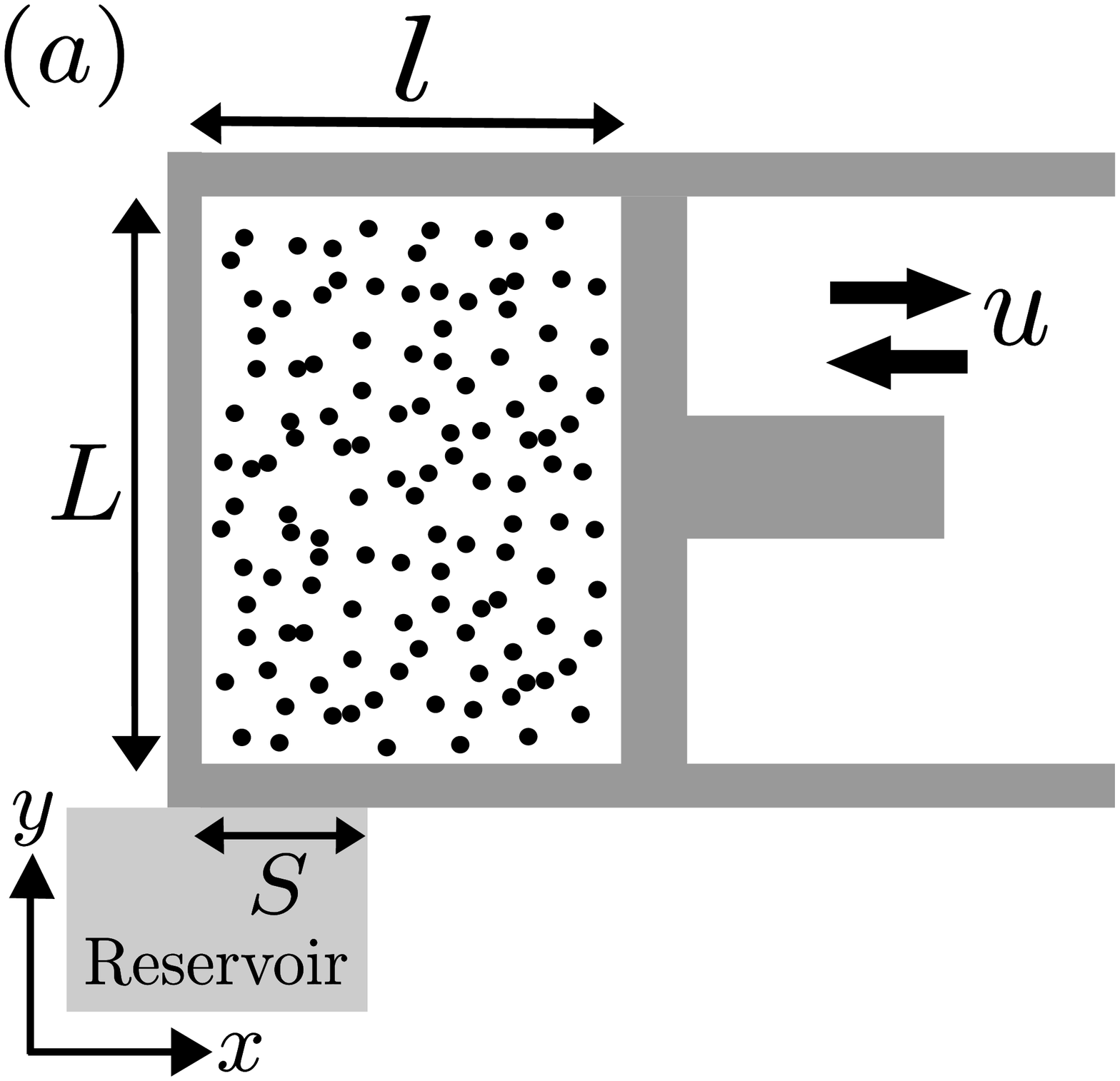}
\includegraphics[scale=0.4]{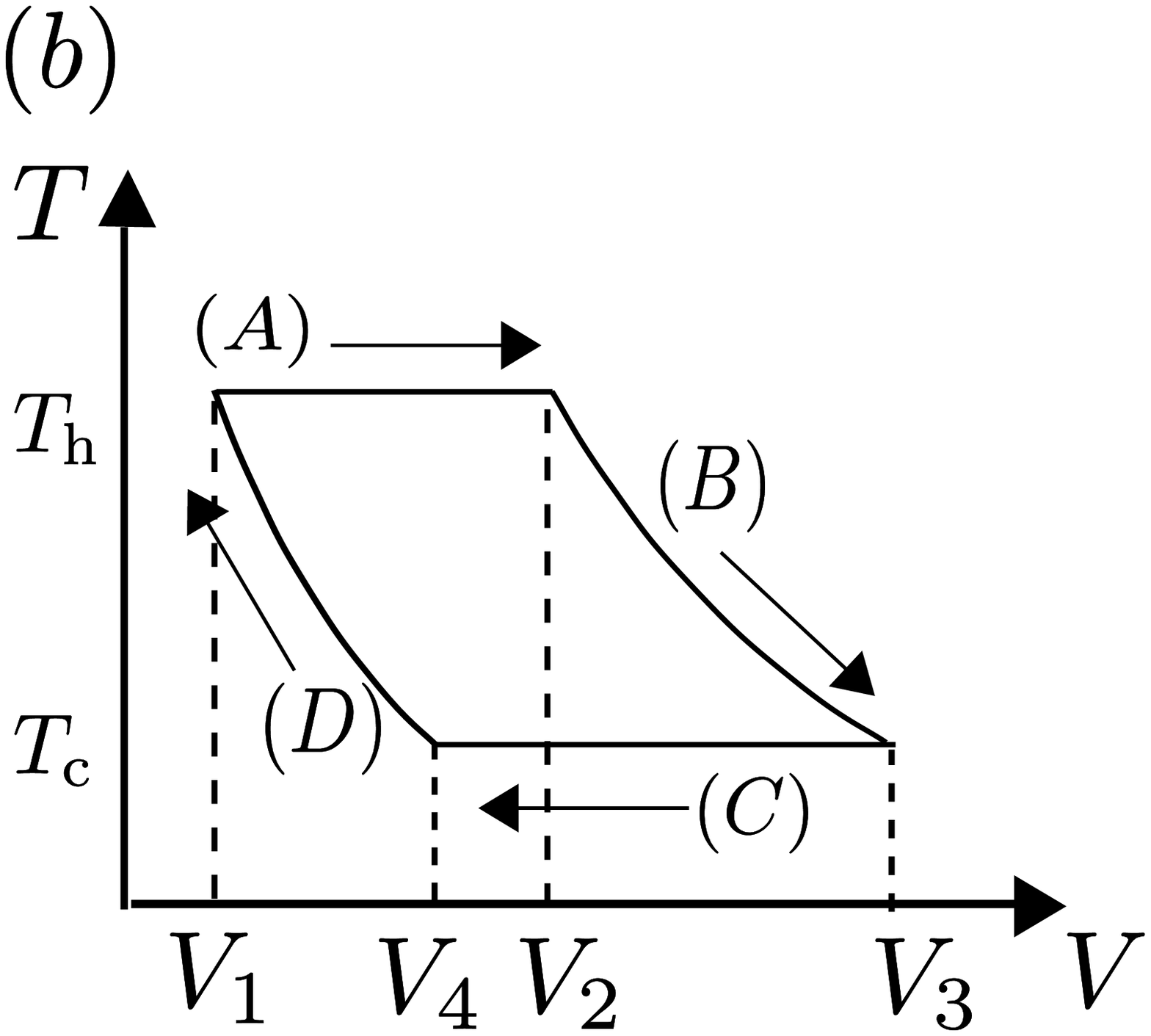}
\end{center}
\caption{(a) Schematic illustration of a two-dimensional finite-time Carnot cycle model. 
Hard-disc particles as a weakly interacting nearly ideal gas are confined into the cylinder.
The piston moves at a finite constant speed $u$ and the thermalizing
wall with
the length $S$ is set on the left bottom of the cylinder during the
isothermal processes. (b) Temperature-volume ($T$-$V$) diagram of a
quasistatic Carnot cycle for a two-dimensional ideal gas.}\label{fig.1}
\end{figure} 
To mimic the weakly interacting nearly ideal gas, 
we confine a low dense
$N$ hard-disc particles with diameter $d$ and mass $m$ into a 
cylinder with rectangular geometry and let them collide with each other. (Fig.~\ref{fig.1} (a)) 
The head of the cylinder is a piston and it moves back and forth at a
constant speed $u$. The usual quasistatic Carnot cycle ($u\to 0$) consists of four
processes (Fig.~\ref{fig.1} (b)): (A): isothermal expansion process ($V_1\to V_2$) in contact
with the hot reservoir at the temperature $T_\mathrm{h}$, (B):
adiabatic expansion process ($V_2\to V_3$), (C): isothermal compression
process ($V_3\to V_4$) in contact with the cold reservoir at
$T_\mathrm{c}$, (D): adiabatic compression process ($V_4\to
V_1$), where $V_k$'s ($k=1,\cdots,4$) are the volumes of the cylinder at which we switch 
each of the four processes. $V_k$'s satisfy the relations 
$V_3=(T_\mathrm{h}/T_\mathrm{c})V_2$ and
$V_4=(T_\mathrm{h}/T_\mathrm{c})V_1$ in the case of the two dimensional
ideal gas. 
Since we regard our system of hard-disc particles as a nearly ideal gas, 
we apply these relations to our system.
In the case of a finite-time cycle, 
we also switch each process at the same volume $V_k$ as in the quasistatic case.

Defining $(x,y)$ coordinates as in Fig.~\ref{fig.1} (a), we let the piston move 
along the $x$-axis at a finite constant speed $u$.
Here, we express the $x$-length and the $y$-length of the
cylinder as $l$ and $L$, respectively and the volume of the cylinder as $V=Ll$. Then, 
the $x$-length $l_k$ at the switching volume $V_k$ can be defined as
$l_k\equiv V_k/L$.

When a particle with the velocity $\mbox{\boldmath$v$}=(v_x, v_y)$
collides with the piston moving at the $x$-velocity $\pm u$, its velocity
changes to $\mbox{\boldmath$v$}^{\prime}=(-v_x\pm 2u,v_y)$, assuming
perfectly elastic collision. Then the
colliding particle gives the microscopic work $m(|\mbox{\boldmath$v$}|^2-|\mbox{\boldmath$v$}^{\prime}|^2)/2=2m(\pm u v_x-u^2)$ against the piston.     
In the isothermal processes (A) and (C), we set the thermalizing wall with
the length $S$ at the position as in Fig.~\ref{fig.1} (a). 
When a particle
with the velocity $\mbox{\boldmath$v$}=(v_x, v_y)$ 
collides with the thermalizing wall, its velocity stochastically changes
to the value governed by the distribution function
\begin{eqnarray}
f(\mbox{\boldmath$v$},T_i)=\frac{1}{\sqrt{2\pi}}\left(\frac{m}{k_\mathrm{B} T_i} \right)^{3/2}v_y \exp\left({-\frac{m\mbox{\boldmath$v$}^2}{2k_\mathrm{B} T_i}}\right)\label{eq.4}
\end{eqnarray}
($-\infty < v_x < +\infty, 0 < v_y
< +\infty$, $T_i$ ($i=\mathrm{h}$ in (A), $\mathrm{c}$ in (C))), where
$k_\mathrm{B}$ is the Boltzmann constant~\cite{TTKB}. The microscopic 
heat flowing from the thermalizing wall can be calculated
by the difference between the kinetic energies before and after the
collision.
We sum up the above microscopic work and heat during one cycle. 
When $Q_\mathrm{h}$ denotes the heat flown
from the hot reservoir during (A) and $Q_\mathrm{c}$ denotes the heat flown from
the cold reservoir during (C), the total work $W$ 
during one cycle is expressed as $W\equiv Q_\mathrm{h}+Q_\mathrm{c}$. Then the
power $P$ and the efficiency $\eta$ can be defined as
$P\equiv \dot{W}$ and $\eta \equiv
W/Q_\mathrm{h}\equiv P/\dot{Q}_\mathrm{h}$.
Here, the dot denotes the value divided by one cycle period or the value per
unit time throughout the paper. In our system,
one cycle period is $2(l_3-l_1)/u$.
\section{Molecular Kinetic Theory}
In this section, we review the results of the molecular kinetic theory
of the finite-time Carnot cycle obtained in our previous work. The 
details of the derivation of the equations in this section are described 
in~\cite{IO}.

If we assume that, even in a finite-time cycle, the gas relaxes to a
uniform equilibrium state with a well-defined temperature $T$ sufficiently fast
and the particle velocity $\mbox{\boldmath$v$}$ is governed by
Maxwell-Boltzmann distribution at $T$, 
we can easily derive the time-evolution equation of $T$ using the 
elementary molecular kinetic theory. Such an assumption of 
the fast relaxation to a uniform equilibrium state at a well-defined
temperature $T$ is valid when the energy equilibration in the cylinder due to the inter-particle
collisions is much faster than the speed of the energy transfers through
the thermalizing wall and the piston.
This situation is surely realized when $u$ is small and the interaction length $S$ between the gas
and the reservoir is sufficiently small. 
If we assume that the gas is sufficiently close to a two-dimensional
ideal gas, the internal energy
of the gas can be approximated as $Nk_\mathrm{B}T$. Then, we can derive
the time-evolution equation of
the gas temperature $T$ for each
of the four processes (A)-(D) as
\begin{eqnarray}
\begin{array}{c c}
(\mathrm{A}): Nk_\mathrm{B} \displaystyle{\frac{dT}{dt}}=q_\mathrm{h}-w_\mathrm{e},&(\mathrm{B}):
Nk_\mathrm{B} \displaystyle{\frac{dT}{dt}}=-w_\mathrm{e},\\
\\
(\mathrm{C}): Nk_\mathrm{B} \displaystyle{\frac{dT}{dt}}=q_\mathrm{c}-w_\mathrm{c},&(\mathrm{D}): Nk_\mathrm{B} \displaystyle{\frac{dT}{dt}}=-w_\mathrm{c},
\end{array}\label{eq.5}
\end{eqnarray}
where $q_i=q_i(t,T)$ ($i=\mathrm{h}$ in
(A), $\mathrm{c}$ in (C)) is the heat flowing into the
system per unit time in the isothermal processes and $w_j=w_j(t,T)$
($j=\mathrm{e}$ in the expansion 
processes (A) and (B),
$\mathrm{c}$ in the compression processes (C)
and (D)) is the    
work against the piston per unit time.
By counting the number of the particles colliding with the thermalizing
wall and the piston, we can derive the specific forms of
$q_i$ and $w_\mathrm{e}$ as
\begin{eqnarray}
&&q_i(t,T)=\frac{3SNk_\mathrm{B}(T_i-T)}{4\pi
 V(t)}\sqrt{\frac{2 \pi k_\mathrm{B}T}{m}},\label{eq.6}
\end{eqnarray}
\begin{eqnarray}
w_\mathrm{e}(t,T)
&&= \frac{2muNL}{V(t)}
\biggl\{\frac{A^2T}{4}-A\sqrt{\frac{T}{\pi}}u+\frac{u^2}{2}\nonumber \\
&&-\int_{0}^{\frac{u}{A\sqrt{T}}}d{v_x} \left(A\sqrt{T}{v_x}-u \right)^2 
\frac{\mathrm{e}^{-{v_x}^2 }}{\sqrt{\pi}}\biggr\},\label{eq.7}
\end{eqnarray}
where $A\equiv
\sqrt{{2k_\mathrm{B}}/{m}}$. $w_\mathrm{c}$ is also
obtained by changing $u \to -u$ in Eq.~(\ref{eq.7}). 
\begin{figure}
\begin{center}
\includegraphics[scale=0.8]{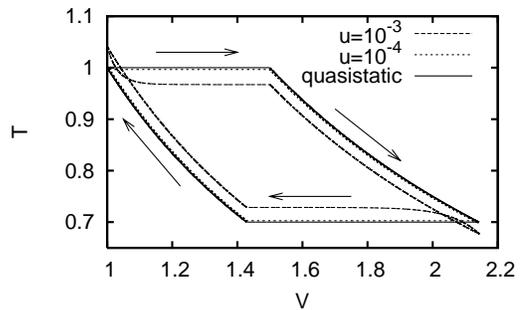}
\end{center}
\caption{Temperature-volume ($T$-$V$) diagram for the steady cycle
obtained by numerically solving Eqs.~(\ref{eq.5}) for $u=10^{-4}$ (dotted line) and
$u=10^{-3}$ (dashed line). The solid line is a theoretical quasistatic
Carnot cycle for a two-dimensional ideal gas. 
Since the dotted line is very close to the solid line, the case of
$u=10^{-4}$ may be regarded as the quasistatic cycle.
The parameters used are $N=100$, 
$T_\mathrm{h}=1$, $T_\mathrm{c}=0.7$, $S=0.05$, $m=1$, $k_\mathrm{B}=1$,
$l_1=1$, $l_2=1.5$ and $L=1$.}\label{fig.2}
\end{figure}
We can numerically solve Eqs.~(\ref{eq.5}) for the entire
cycle at various piston speeds. By using the final temperature of each process as the initial
temperature of the next process repeatedly, we can obtain a steady
cycle at a given $u$. In Fig.~\ref{fig.2}, we have shown the temperature-volume ($T$-$V$) diagram for the steady
cycle at $u=10^{-4}$ (dotted line) and $u=10^{-3}$ (dashed line), respectively. The solid line is the theoretical
quasistatic line of a two-dimensional ideal gas. From this
figure, we can see that as $u$ becomes
larger, the cycle deviates from the theoretical quasistatic line and
the temperatures during the isothermal processes (A) and (C) relax to the
steady temperatures $T_\mathrm{h}^{\mathrm{st}} (<T_\mathrm{h})$ and
$T_\mathrm{c}^{\mathrm{st}} (>T_\mathrm{c})$, respectively. 
If it is considered that the relaxation to the steady temperature is
very fast, the isothermal process may approximately be divided into the 
relaxational part and the steady part in the case of a finite-time
cycle, where $T(t)$ instantaneously changes from the initial
temperature of the isothermal process to the steady temperature in the
relaxational part and $T(t)$ keeps the steady temperature in the steady
part. These relaxational parts in the isothermal processes (A) and (C)
are surely missed in the original model by Curzon and
Ahlborn~\cite{CA,IO,BJM}.
Moreover we note that the heat flow per unit time $q_i$ as shown
in Eq.~(\ref{eq.6}) is time-dependent even during $T=T_i^{\mathrm{st}}$ 
through the volume of the cylinder after the relaxational processes in
(A) and (C). Therefore the assumption of the time-independent heat flows
in the original phenomenological model in~\cite{CA} does not seem to be
applied to our model. 

Although it may be
impossible to find the exact solutions of Eqs.~(\ref{eq.5}), 
we can find the analytic expressions of the steady temperature 
$T_i^\mathrm{st}$ during the isothermal processes (A) and (C) as the 
solutions of $dT/dt=0$ by expanding as $T_i^\mathrm{st}=T_i+a_i^1 u+a_i^2
u^2+O(u^3)$, assuming that $u$ is small. These expansion coefficients $a_i^1,
a_i^2$, etc. can be
determined order by order. If we assume that the relaxational process to
$T_\mathrm{h}^{\mathrm{st}}$ is
sufficiently fast, the heat $Q_\mathrm{h}^\mathrm{st}$ flowing into the system during
the steady part $T(t)=T_\mathrm{h}^\mathrm{st}$ can be calculated up to $O(u)$ as:
\begin{eqnarray}
Q_\mathrm{h}^\mathrm{st}&&=\int_{0}^{(l_2-l_1)/u}
q_\mathrm{h}(t,T_\mathrm{h}^{\mathrm{st}})\ dt \nonumber \\
&&=Q_\mathrm{h}^{\mathrm{qs}}-2mNA\sqrt{T_\mathrm{h} \pi}\left(\frac{1}{\pi}+\frac{L}{3S}\right)u
\ln\frac{V_2}{V_1},\label{eq.8}
\end{eqnarray}
where the quasistatic heat for an ideal gas in the isothermal process (A)
is defined as $Q_\mathrm{h}^\mathrm{qs}=Nk_\mathrm{B}T_\mathrm{h}\ln
(V_2/V_1)$. $Q_\mathrm{c}^\mathrm{st}$ during
$T(t)=T_\mathrm{c}^{\mathrm{st}}$
can also be obtained by changing 
$u \to -u$, $T_\mathrm{h}\to T_\mathrm{c}$ and $V_2/V_1 \to V_4/V_3$ in Eq.~(\ref{eq.8}).
If we neglect the contributions of the heat transfers
through the thermalizing wall during the relaxational parts in (A) and (C),
the steady part of the work during one cycle $W^\mathrm{st}\equiv
Q_\mathrm{h}^\mathrm{st}+Q_\mathrm{c}^\mathrm{st}$ becomes
\begin{eqnarray}
W^\mathrm{st}&&=Nk_\mathrm{B}(T_\mathrm{h}-T_\mathrm{c})\ln\frac{V_2}{V_1}-2mNA\sqrt{\pi}\left(\frac{1}{\pi}+\frac{L}{3S}\right)\nonumber\\
&&\times (\sqrt{T_\mathrm{h}}+\sqrt{T_\mathrm{c}})u\ln\frac{V_2}{V_1}.\label{eq.9}
\end{eqnarray} 
If we assume that the adiabatic processes (B) and (D) even at a finite
$u$ satisfy the quasistatic adiabatic relation $TV=\mathrm{const}$ of a
two-dimensional ideal gas,                                      
the additional heat transfers $Q_\mathrm{h}^{\mathrm{add}}$ and
$Q_\mathrm{c}^{\mathrm{add}}$ during the relaxational part in (A) and
(C), respectively, can be estimated as 
\begin{eqnarray} 
&&Q_\mathrm{h}^{\mathrm{add}}=-Nk_\mathrm{B} \frac{4L\sqrt{\pi
T_\mathrm{h}}}{3SA}\left(1+\sqrt{\frac{T_\mathrm{h}}{T_\mathrm{c}}}\right)u
\label{eq.10} 
\end{eqnarray}     
up to $O(u)$ and $Q_\mathrm{c}^{\mathrm{add}}$ can also be
obtained by changing $u \to -u$ and $T_\mathrm{h} \leftrightarrow
T_\mathrm{c}$ in Eq.~(\ref{eq.10}). Therefore the total $Q_\mathrm{h}$,
$Q_\mathrm{c}$ and $W$ become
$Q_\mathrm{h}=Q_\mathrm{h}^\mathrm{st}+Q_\mathrm{h}^\mathrm{add}$,
$Q_\mathrm{c}=Q_\mathrm{c}^\mathrm{st}+Q_\mathrm{c}^\mathrm{add}$ and
$W=W^\mathrm{st}+Q_\mathrm{h}^\mathrm{add}+Q_\mathrm{c}^\mathrm{add}$,
respectively. 
\section{Calculation of Onsager coefficients}
After the preliminaries in the previous sections, we can now calculate the Onsager coefficients
for our finite-time Carnot cycle in the linear-response regime
$\Delta T \to 0$ as follows. The first step is an appropriate choice of the
thermodynamic forces and fluxes for this system. A typical way to choose
these thermodynamic forces and fluxes is to introduce  
the rate of the total entropy production $\dot{\sigma}$ during one cycle: 
\begin{eqnarray}
\dot{\sigma}\equiv -\frac{\dot{Q}_\mathrm{h}}{T_\mathrm{h}}-\frac{\dot{Q}_\mathrm{c}}{T_\mathrm{c}}.\label{eq.11}
\end{eqnarray}
This is just the entropy increase of the two reservoirs during one cycle
divided by the 
cycle period $2(l_3-l_1)/u$
because the entropy of the working substance does not change after one cycle. 
Using the relation $Q_\mathrm{c}=W-Q_\mathrm{h}$ and considering the 
linear-response regime $\Delta T \to 0$, it can be rewritten as 
\begin{eqnarray}
\dot{\sigma}=\frac{u(-W)}{2(l_3-l_1)T}+\frac{\Delta T}{T^2}\dot{Q}_\mathrm{h},\label{eq.12}
\end{eqnarray}
where $T\equiv (T_\mathrm{h}+T_\mathrm{c})/2$ and
we have neglected ${\Delta T}^3\dot{Q}_\mathrm{h}$ and $uW\Delta T$ terms,  
the reason of which we will clarify later. According to the
linear-response theory,
$\dot{\sigma}$ can be expressed as the sum of the product of the thermodynamic
force and its conjugate thermodynamic flux~\cite{O,GM}:
\begin{eqnarray}
\dot{\sigma}=J_1X_1+J_2X_2,
\label{eq.13}
\end{eqnarray}
where we define the thermodynamic forces as
\begin{eqnarray}
X_1 \equiv \frac{-W}{2(l_3-l_1)T},\ X_2 \equiv \frac{\Delta T}{T^2}\label{eq.14}
\end{eqnarray}
and their conjugate fluxes as
\begin{eqnarray}
J_1 \equiv u,\ J_2 \equiv \dot{Q}_\mathrm{h}.\label{eq.15}
\end{eqnarray}
Moreover the linear-response theory assumes the Onsager relations between the fluxes and forces~\cite{O,GM}: 
\begin{eqnarray}
J_1&&=u=L_{11}\frac{-W}{2(l_3-l_1)T}+L_{12}\frac{\Delta T}{T^2}, \label{eq.16}\\
J_2&&=\dot{Q}_\mathrm{h}=L_{21}\frac{-W}{2(l_3-l_1)T}+L_{22}\frac{\Delta T}{T^2},\label{eq.17}
\end{eqnarray} 
where $L_{ij}$'s are the Onsager coefficients and the non-diagonal
elements should satisfy the symmetry relation $L_{12}=L_{21}$. 
From these relations between the thermodynamic fluxes
and forces, we understand that $\dot{\sigma}=J_1X_1+J_2X_2$ is the quantity of the
second-order of the thermodynamic forces, 
which explain why we neglected the higher order terms like ${\Delta T}^3\dot{Q}_\mathrm{h}$ 
and $uW\Delta T$ in Eq.~(\ref{eq.12}). Moreover, although we considered
the contributions of the additional heat transfers $Q_\mathrm{h}^\mathrm{add}$ and
$Q_\mathrm{c}^\mathrm{add}$ to the total heat and the work in
Sec.~III, 
we can easily show that their effects
do not contribute to $\dot{\sigma}$ in the limit of $\Delta T \to 0$.
Therefore we can indeed neglect
them in the linear-response regime and use $Q_\mathrm{h}=Q_\mathrm{h}^\mathrm{st}$ and $W=W^\mathrm{st}$ 
in the calculations of the Onsager coefficients below. 

We are now in a position to calculate the Onsager coefficients
explicitly. First we determine $L_{11}$ and $L_{21}$ as follows. To calculate 
$L_{11}$, we consider the relation between $u$ and 
$X_1$ in the case of $\Delta T=0$. 
Expanding Eq.~(\ref{eq.9}) by
$\Delta T$ and putting $\Delta T=0$, we can obtain the relation
\begin{eqnarray}
W=-4mNA\sqrt{\pi
T}\left(\frac{1}{\pi}+\frac{L}{3S}\right)\ln\frac{V_2}{V_1}\times u.\label{eq.18}
\end{eqnarray}
Comparing Eq.~(\ref{eq.18}) with Eq.~(\ref{eq.16}), $L_{11}$ is determined as 
\begin{eqnarray}
L_{11}=\frac{(l_3-l_1)T^{1/2}}{2mN(\frac{1}{\pi}+\frac{L}{3S})\sqrt{\frac{2\pi
 k_\mathrm{B}}{m}}\ln \frac{V_2}{V_1}}. \label{eq.19}
\end{eqnarray}
Likewise $\dot{Q}_\mathrm{h}$ with $\Delta T=0$ can be evaluated up to
the linear order in $W$ using Eq.~(\ref{eq.8}) and (\ref{eq.18}) as
\begin{eqnarray}
\dot{Q}_\mathrm{h}=\frac{k_\mathrm{B}T^{3/2}}{4m(\frac{1}{\pi}+\frac{L}{3S})\sqrt{\frac{2\pi
k_\mathrm{B}}{m}}}\frac{-W}{2(l_3-l_1)T}.\label{eq.20}
\end{eqnarray}
Comparing Eq.~(\ref{eq.20}) with Eq.~(\ref{eq.17}), $L_{21}$ is
determined as 
\begin{eqnarray}
L_{21}=\frac{k_\mathrm{B}T^{3/2}}{4m(\frac{1}{\pi}+\frac{L}{3S})\sqrt{\frac{2\pi
k_\mathrm{B}}{m}}}.\label{eq.21}
\end{eqnarray}
Next we determine $L_{12}$ and $L_{22}$ as follows. To calculate
$L_{12}$, we consider the relation between $u$ and $X_2$ in the
case of $W=0$. In this case, regardless of a finite temperature
difference, useful work cannot be obtained because the
engine runs so fast that it cannot output positive work. 
Therefore, this case is called the work-consuming state. The speed of
the piston at the work-consuming state 
can be obtained as a solution of $W=0$ in
Eq.~(\ref{eq.9}). Considering in the linear-response regime 
$\sqrt{T_\mathrm{h}}-\sqrt{T_\mathrm{c}}\simeq \Delta T/(2\sqrt{T})$, it becomes
\begin{eqnarray}
u&&=\frac{k_\mathrm{B} T^{3/2}}{4m(\frac{1}{\pi}+\frac{L}{3S})\sqrt{\frac{2\pi
k_\mathrm{B}}{m}}}\frac{\Delta T}{T^2}.\label{eq.22}
\end{eqnarray}  
From Eq.~(\ref{eq.16}) and Eq.~(\ref{eq.22}), we can obtain
\begin{eqnarray}
L_{12}=\frac{k_\mathrm{B}T^{3/2}}{4m(\frac{1}{\pi}+\frac{L}{3S})\sqrt{\frac{2\pi
k_\mathrm{B}}{m}}}.\label{eq.23}
\end{eqnarray}
From Eqs.~(\ref{eq.21}) and (\ref{eq.23}), we can confirm the symmetry relation
$L_{12}=L_{21}$ as expected. 
To determine the last coefficient $L_{22}$, 
we consider the heat flow $\dot{Q}_\mathrm{h}$ at the work-consuming
state using Eq.~(\ref{eq.8}) and (\ref{eq.22}):
\begin{eqnarray}
\dot{Q}_\mathrm{h}=\frac{Nk_\mathrm{B}^2\ln\frac{V_2}{V_1}}{8m(\frac{1}{\pi}+\frac{L}{3S})\sqrt{\frac{2\pi
_\mathrm{B}}{m}}(l_3-l_1)}T^{5/2}\frac{\Delta T}{T^2}.\label{eq.24}
\end{eqnarray}
From Eq.~(\ref{eq.17}) and Eq.~(\ref{eq.24}), the last coefficient $L_{22}$ turns out to be
\begin{eqnarray}
L_{22}=\frac{Nk_\mathrm{B}^2\ln\frac{V_2}{V_1}T^{5/2}}{8m(\frac{1}{\pi}+\frac{L}{3S})\sqrt{\frac{2\pi
_\mathrm{B}}{m}}(l_3-l_1)}.\label{eq.25}
\end{eqnarray}
Note that the positivity of
the rate of the total entropy production, $\dot{\sigma}=J_1 X_1+J_2 X_2\ge 0$, 
should restrict the values of the Onsager coefficients to $L_{11}\ge 0$, $L_{22}\ge 0$ 
and $L_{11}L_{22}-L_{12}L_{21}\ge 0$. We can confirm that $L_{ij}$ of our finite-time Carnot
cycle surely satisfy these relations.
These analytic expressions of the Onsager coefficients $L_{ij}$ 
are the main result of this paper.

To confirm the validity of the above analytic
calculations of the Onsager coefficients, especially $T$
dependence of them, we performed the event-driven molecular dynamics (MD) computer simulations~\cite{IO,IO2,AW} of our
two-dimensional finite-time Carnot cycle, following the procedure described in Sec.~II.

\begin{figure}
\begin{center}
\includegraphics[scale=1.55]{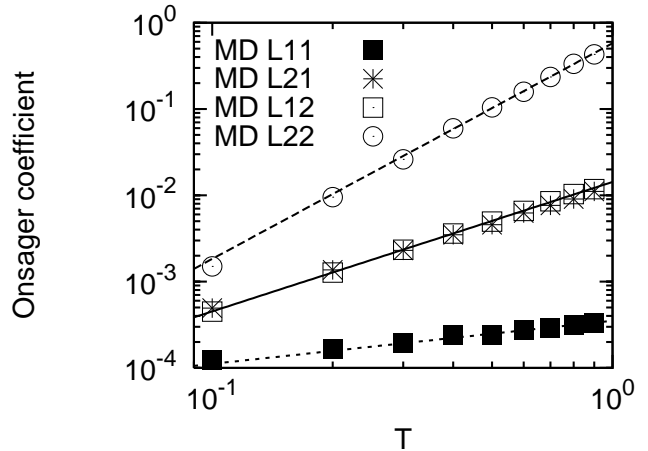}
\end{center}
\caption{$T$-dependence of the Onsager coefficients.
The dashed line, the solid line and the dotted line indicate the
theoretical Onsager coefficient $L_{22}$ (Eq.~(\ref{eq.25})), $L_{12}$
(Eq.~(\ref{eq.23})) ($=L_{21}$ (Eq.~(\ref{eq.21}))) and 
$L_{11}$ (Eq.~(\ref{eq.19})), respectively. The parameters used in the MD
simulations are $\Delta T=4\times 10^{-3}$ and $d=0.01$. The other
parameters are the same as in Fig.~\ref{fig.2}. 
The MD data were obtained by averaging $100$-$800$  
cycles after transient $5$ cycles in the simulations.}\label{fig.3}
\end{figure}

To calculate $L_{12}$ and $L_{22}$ at given $T$, 
we fix the temperature difference $\Delta T$ to a sufficiently small value
and find the piston speed where the work becomes $0$. 
Then we can determine $L_{12}$
and $L_{22}$ numerically as 
\begin{eqnarray}
L_{12}&&=\frac{J_1}{X_2}=\frac{uT^2}{\Delta T},\label{eq.26}\\
L_{22}&&=\frac{J_2}{X_2}=\frac{\dot{Q}_\mathrm{h}T^2}{\Delta T}\label{eq.27}
\end{eqnarray}
from Eq.~(\ref{eq.16})
and Eq.~(\ref{eq.17}). Next, to calculate $L_{11}$ and 
$L_{21}$ at given $T$, we set $\Delta T=0$. Fixing $u$ to a
sufficiently small value
($2.5\times 10^{-5}\sim  3\times 10^{-4}$), we can determine
$L_{11}$ and $L_{21}$ numerically as 
\begin{eqnarray}
L_{11}&&=\frac{J_1}{X_1}=-\frac{2(l_3-l_1)Tu}{W},\label{eq.28}\\
L_{21}&&=\frac{J_2}{X_1}=-\frac{2(l_3-l_1)T\dot{Q}_\mathrm{h}}{W}\label{eq.29}
\end{eqnarray}
from using
Eq.~(\ref{eq.16}) and Eq.~(\ref{eq.17}). Fig.~\ref{fig.3} shows
$T$ dependence of these Onsager coefficients determined by the MD
simulations as
well as the
analytic results. We can see fairly good agreement
between the MD data and the analytic lines in the range we studied. These
data clearly support the validity of our analytic expressions of the Onsager
coefficients which has been obtained under some theoretical
assumptions.  
\section{The efficiency at the maximal power}
To see how the Onsager coefficients derived in Sec.~IV indeed govern the
behavior of the finite-time Carnot cycle, 
we briefly introduce the general framework of the heat engine 
obeying the Onsager relations~\cite{VB}.

\begin{figure}[t]
\begin{center}
\includegraphics[scale=0.4]{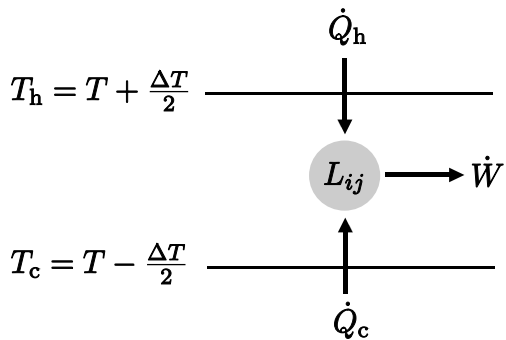}
\end{center}
\caption{Schematic illustration of the heat engine governed by the Onsager
relations Eq.~(\ref{eq.30}) and Eq.~(\ref{eq.31}).}\label{fig.4}
\end{figure}

The basic setup is as follows.
We consider a general steady or cyclic process in which the work is extracted 
from the heat flow between the small temperature difference. (see Fig.~\ref{fig.4})
The work $W$ done against the external force $F$ is
$W=-Fx$ where $x$ is the thermodynamically conjugate variable of $F$.  
We define a thermodynamic force as $X_1=F/T$ and the
corresponding thermodynamic flux as
$J_1=\dot{x}$. We also choose
$X_2=1/T_\mathrm{c}-1/T_\mathrm{h}$ as another thermodynamic force and
$J_2=\dot{Q}_\mathrm{h}$ as the corresponding thermodynamic flux. 
If we consider the linear-response regime $\Delta T \to 0$,
$X_2$ can be written as $X_2\simeq \Delta T/T^2$.
Moreover the linear-response theory assumes the Onsager
relations between the thermodynamic forces and fluxes~\cite{O,GM}:
\begin{eqnarray}
J_1&&=L_{11}X_1+L_{12}X_2, \label{eq.30} \\
J_2&&=L_{21}X_1+L_{22}X_2,\label{eq.31}
\end{eqnarray} 
where the non-diagonal elements of the Onsager
coefficients should satisfy $L_{12}=L_{21}$. 
Then, the power $P=\dot{W}$ and the efficiency $\eta=P/\dot{Q}_\mathrm{h}$ of the engine can be expressed as 
\begin{eqnarray}
P&&=-J_1X_1T,\label{eq.32}\\
\eta&&=\frac{-J_1X_1T}{J_2}.\label{eq.33}
\end{eqnarray}
Now the efficiency at the maximal power $\eta_\mathrm{max}$ can be given as follows:
We first maximize the power at $X_1=-L_{12}X_2/(2L_{11})$ which is 
determined as the solution of 
$\partial P/\partial X_1=0$,
then, $\eta_\mathrm{max}$ becomes
\begin{eqnarray}
\eta_\mathrm{max}=\frac{\Delta T}{2T}\frac{q^2}{2-q^2},\label{eq.34}
\end{eqnarray}
where 
\begin{eqnarray}
q\equiv \frac{L_{12}}{\sqrt{L_{11}L_{22}}}\label{eq.35}
\end{eqnarray} 
is called the coupling strength parameter.
Note that it takes $-1 \le q \le +1$ due to the positivity of $\dot{\sigma}$.
When $q$ satisfies the tight-coupling condition $|q|=1$,
$\eta_\mathrm{max}$ takes the maximal value
$\Delta T/(2T)$, which is equal to the CA efficiency up to
the lowest order in $\Delta T$. 

In Refs.~\cite{IO,IO2}, we studied $\eta_\mathrm{max}$ of this finite-time Carnot cycle 
extensively by performing the MD computer simulations as a numerical experiment to
verify the validity of the CA efficiency. We found there that $\eta_\mathrm{max}$ agrees with
the CA efficiency in the limit of $\Delta T \to 0$. Our molecular
kinetic theory also confirmed this property~\cite{IO}: Using 
$Q_i=Q_i^\mathrm{st}+Q_i^\mathrm{add}$ in Sec.~III, we can calculate the 
speed $u=u_\mathrm{max}$ at which the power $P$ maximizes as
\begin{eqnarray}
u_{\mathrm{max}}&&=\frac{k_\mathrm{B} (T_\mathrm{h}-T_\mathrm{c})} 
{\sqrt{\pi}}\Biggl\{4mA\left(\frac{L}{3S}+\frac{1}{\pi}\right) \nonumber \\
&&\times
(\sqrt{T_\mathrm{h}}+\sqrt{T_\mathrm{c}})\ln \frac{V_2}{V_1}+\frac{8L
k_\mathrm{B}}{3SA\sqrt{T_\mathrm{h}}}\nonumber\\
&&\times \left(T_\mathrm{h}-T_\mathrm{c}\right)\left(1+\sqrt{\frac{T_\mathrm{h}}{T_\mathrm{c}}}\right)\Biggr\}^{-1}\ln
\frac{V_2}{V_1},\label{eq.36}
\end{eqnarray}
and can confirm that the efficiency $\eta$ at $u=u_\mathrm{max}$ shows
\begin{eqnarray}
\eta_\mathrm{max}\equiv
 \eta(u_\mathrm{max})&&=\frac{W(u_\mathrm{max})}{Q_\mathrm{h}^{\mathrm{st}}(u_{\mathrm{max}})
+Q_\mathrm{h}^{\mathrm{add}}(u_{\mathrm{max}})}\nonumber \\
&&\to \frac{\Delta T}{2T} \ (\Delta T \to 0).\label{eq.37}
\end{eqnarray}
Now we can clarify
the underlying physics of this behavior of $\eta_\mathrm{max}$.
In the limit of $\Delta T \to 0$, 
our finite-time Carnot cycle can be described by the
Onsager relations as shown in Sec~IV. Then we can confirm that $q=1$ in our 
finite-time Carnot cycle
from Eqs.~(\ref{eq.19}), (\ref{eq.23}), (\ref{eq.25}) and
(\ref{eq.35}). 
This condition gives a proof 
that our finite-time Carnot cycle shows the CA
efficiency in the limit of $\Delta T \to 0$ as suggested in~\cite{IO,IO2}
from the viewpoint of the linear-response theory. 
\section{Summary}
In this paper, we have studied a finite-time Carnot cycle of a
two-dimensional weakly interacting nearly ideal gas working in the
linear-response regime and have explicitly
calculated the Onsager coefficients of this system for the first time.
Molecular dynamics computer simulations of this cycle have supported the theoretical
calculations in spite of some assumptions in the analysis. 
We have revealed that the Onsager coefficients of this system satisfy 
the tight-coupling condition $q=L_{12}/\sqrt{L_{11}L_{22}}=1$ and
therefore can understand why our finite-time Carnot cycle attains the
Curzon-Ahlborn efficiency in the linear-response regime $\Delta T \to
0$ as suggested in Refs.~\cite{IO,IO2}.
It would be an interesting problem to construct and study the
collective behavior of the Carnot
cycles coupled with each other by using the property of the Onsager coefficients derived in
this paper~\cite{OAMB,VB,VB2,JA,JA2}.
\begin{acknowledgements} 
The authors thank M. Hoshina for helpful discussions.
This work was supported by the 21st Century Center of Excellence (COE)
program entitled ``Topological Science and Technology'', Hokkaido University.
\end{acknowledgements}

\bibliography{basename of .bib file}

\end{document}